\documentclass[pra,preprint]{revtex4}
\usepackage{graphicx}
\usepackage{natbib}

\begin{document}
\title{Squeezing as the source of inefficiency in the quantum Otto cycle}
\author{A.M. Zagoskin$^{1,2}$, S. Savel'ev$^{1,2}$, Franco Nori$^{2,3}$, and F.V. Kusmartsev$^{1}$}

\affiliation{$^{1}$  Department of Physics, Loughborough University, Loughborough LE11 3TU, United Kingdom \\  $^{2}$   Advanced Science Institute, RIKEN, Wako-shi, Saitama, 351-0198, Japan\\ $^{3}$ Department of Physics,  University of Michigan, Ann Arbor, MI 48109-1040, USA }

\begin{abstract}
The availability of controllable macroscopic devices, which maintain quantum coherence over relatively long time intervals, for the first time allows an experimental realization of many effects previously considered only as Gedankenexperiments, such as the operation of quantum heat engines. The theoretical efficiency $\eta$ of quantum heat engines is restricted by the same Carnot boundary $\eta_C$ as for the classical ones: any deviations from quasistatic evolution suppressing $\eta$ below $\eta_C$. Here we investigate an implementation of an analog of the Otto cycle in a tunable quantum coherent circuit and show that the specific source of inefficiency is the quantum squeezing of the thermal state due to the finite speed of compression/expansion of the system. 
\end{abstract}

\maketitle

\section{Quasistatic quantum Otto cycle}

Quantum heat engines (QHE) are engines with a quantum coherent working body. After they were   introduced \cite{Scovil1959,Geusic1967} as the generalization of classical heat engines for lasing, they were extensively used in Gedankenexperiments helping clarify subtle points of statistical mechanics, like the role of fluctuations and the operation of Maxwell's demon in the quantum regime (see, e.g., \cite{Geva1994,Scully2003,Kieu2004,Humphrey2005,Quan2007,Quan2009,Maruyama2009}).  Even  though QHEs demonstrate some very nonclassical features, of course these do not violate the second law of thermodynamics. In particular, QHEs satisfy the Carnot inequality,
\begin{equation}
\eta = \frac{R}{Q_h}  \leq \eta_C = 1 - \frac{T_c}{T_h},
\label{eq:1-carnot}
\end{equation} 
where $R$ is the mechanical work performed per cycle, $Q_h$ is the energy obtained from the heater (an equilibrium reservoir at temperature $T_h$); and $Q_c$ is the energy transferred to the cooler (at temperature $T_c$). 

The latest developments in   solid state-based quantum computing (see, e.g., \cite{Makhlin2001,Wendin2005a,Zagoskin2007a,Zagoskin2011,You2011,Buluta2011,Hanson2008,Morton2011}), especially with superconducting devices, gave this line of investigation more experimental relevance. Unlike the various mesoscopic cooling schemes in operation since the mid-1990s (see \cite{Giazotto2006}), the sideband cooling of superconducting qubits, achieved in \cite{Valenzuela2006,Grajcar2008}, does not involve the exchange of particles between the system and the thermal reservoirs and is therefore much closer to Gedankenexperiments' QHEs.

Any heat engine operating along any other thermodynamic cycle is less efficient than the ideal Carnot engine. Moreover, any heat engine operating at finite speed is less efficient than the same engine operating quasistatically. 
The source of inefficiency is in   the irreversibility of non-quasistatic processes, but the specific mechanism for the QHE depends on the particular realization and is often assumed to be due to the loss of quantum coherence. In this paper we consider the quantum Otto cycle (see, e.g., \cite{Kieu2004,Feldmann2004,Feldmann2006,Rezek2006,Quan2006,Quan2007,Quan2009,Feldmann2010}) and show that the source  of its inefficiency at finite operation speed (compared to the quasistatic case) lies in the reversible, quantum coherent  squeezing of the quantum state of the working body.

Let us consider a textbook situation, where there are two thermal reservoirs at temperatures $T_c, T_h$, and a system which can be periodically put in thermal equilibrium with either of them (e.g., \cite{Fermi1956}). In its quantum analog, the reservoirs and the system are represented by three harmonic oscillators, with frequencies $\omega_c, \omega_h,$ and $ \omega$, respectively, assuming that $\omega$ can be changed at will. In the following, the nature of these oscillators  is immaterial. For the sake of definitiveness, we will talk about superconducting $LC$-circuits  with   a Josephson junction, which can be biased by a current or an external magnetic flux. The tunability of such a system in the quantum coherent regime was repeatedly demonstrated in  experiments (e.g., \cite{Ploeg2006,Harris2007}). We assume that 
\begin{equation}
\omega_h > \omega_c,
\label{eq:1-ineq}
\end{equation}
but will not impose any restrictions on the baths temperatures. In such an implementation, the mechanical work produced by, or applied to, the system is electrical and is determined by the quantum state-dependent  energy required to  change the bias current in a Josephson junction.

The quantum Otto cycle runs as follows (Fig.~\ref{fig:1}). The system is initially tuned to and equilibrated with  the ``cold" reservoir (A). Then it is quasistatically tuned in resonance with the ``hot" reservoir. According to the quantum adiabatic theorem \cite{Landau2003}, the occupation numbers of its energy eigenstates will not change during this process, while mechanical work must be performed to increase $\omega$. At point (B) the system is in resonance, but out of equilibrium, with the ``hot" reservoir, $\omega = \omega_h$. After the equilibration is achieved (C), the system is quasistatically brought back in resonance with the ``cold" reservoir, and is  again equilibrated with it at (A), closing the cycle.  On the adiabatic stages AB and CD, the system is isolated from the outside world (neglecting the finite, but ideally vanishing linewidth of the system and reservoirs' energy levels), and all the energy exchange takes place along the ``isochoric" stages BC and DA. The work performed by the system is given by the area enclosed by the contour ABCD:
\begin{equation}
R = \int_A^B   \langle n\rangle \: d\omega -  \int_C^D   \langle n\rangle \: d\omega = \pm {\cal A}_{ABCD},
\label{eq:1-work}
\end{equation}
where $\langle n\rangle$ is the expectation value of the photon number in the system. 
The role of volume is played by the inverse frequency, $1/\omega$, and the pressure in the system is $ \langle n \rangle\omega^2$.   
For the clockwise sense $R>0$, and the device works as a heat engine; otherwise $R<0$, and it is a refrigerator, transferring energy from the ``cold" to the ``hot"  reservoir. 

The calculations for the ideal Otto cycle are straightforward: denoting by $|n,c(h)\rangle$ the $n$th energy eigenstate of an oscillator with frequency $\omega_c (\omega_h)$, and by $Z_{c(h)}$ the corresponding partition function, we can write for the density matrix of the system at points A, B, C, D
\begin{eqnarray}
\rho_A = Z_c^{-1}\sum_n  e^{-\frac{ \omega_cn}{T_c}} |n,c\rangle\langle n,c|,\:\:
\rho_B = Z_c^{-1}\sum_n  e^{-\frac{ \omega_cn}{T_c}} |n,h\rangle\langle n,h|,\nonumber\\
\rho_C = Z_h^{-1}\sum_n  e^{-\frac{ \omega_hn}{T_h}} |n,h\rangle\langle n,h|,\:\:
\rho_D = Z_h^{-1}\sum_n  e^{-\frac{ \omega_hn}{T_h}} |n,c\rangle\langle n,c|,
\label{eq:1-density-matrices}
\end{eqnarray} 
which immediately yields the expressions for the energy of the system at the corresponding points and for the work done  and the energy received  by the system:

\begin{figure}%
\includegraphics[width=3.2in]{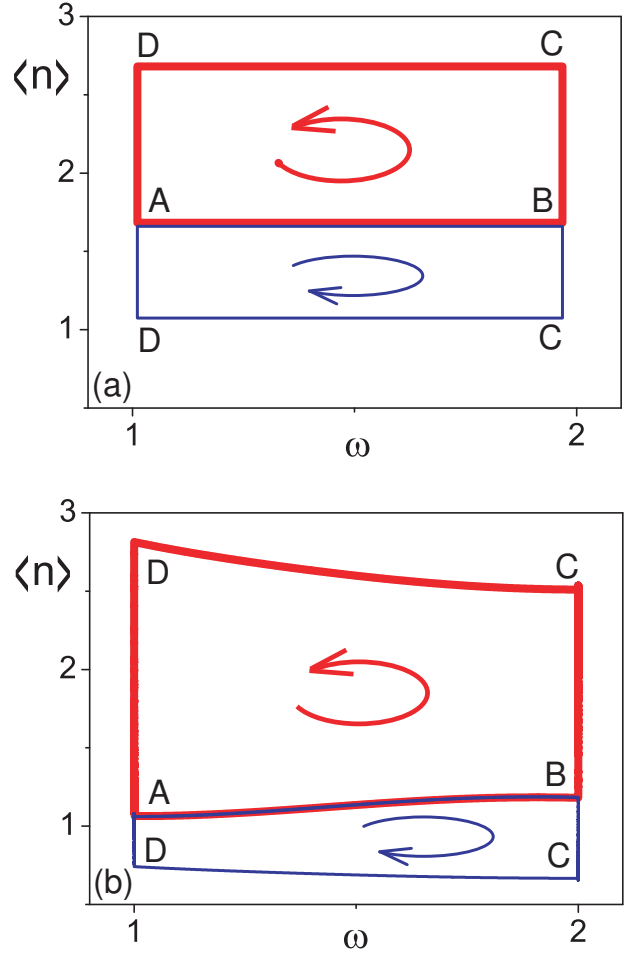}
\caption{(Color online.) The expectation value of the photon number $\langle n \rangle$ in the tunable oscillator (``working body") versus the oscillator frequency $\omega$.  This quantum Otto cycle is for a system comprising of two oscillators in thermal equilibrium (``reservoirs") and a tunable oscillator (``working body"). This cycle is realized by changing the latter's frequency between $\omega_c$ and $\omega_h$ along AB and CD (adiabatic stages). Along BC and DA (isochoric stages), the tunable oscillator equilibrates with the reservoirs. Whether the system operates as a heat engine (red, upper cycle) or a heat pump (blue, lower cycle) is determined by the parameter $\lambda =   \omega_c/T_c - \omega_h/T_h$, Eq.~(\ref{eq:1-parameter}).  
(a) Quasistatic case. (b) Finite speed case [calculated using Eq.~(\ref{eq:5-44bis}) with the oscillator frequency $\omega$ as a function of time shown in the inset of Fig.~\ref{fig-entropy}].}%
\label{fig:1}%
\end{figure}

\begin{eqnarray}
R_1 = E_A - E_B = -{ (\omega_h-\omega_c)}{\nu_c}  ,\:
R_2 = E_C - E_D = { (\omega_h-\omega_c)}{\nu_h}    ,\: \nonumber\\
R = R_1 + R_2 = { (\omega_h-\omega_c)}({\nu_h-\nu_c});  ~~~~~~~~~~~~~~~~~~~~~~ \\
Q_1 = E_C - E_B = { \omega_h} ({\nu_h-\nu_c});\:
Q_2 = E_A - E_D = -{ \omega_c} ({\nu_h-\nu_c}), \nonumber
 \label{eq:1-work-and-heat}
\end{eqnarray}
where $\nu_{i}=\bar{n}(x_i)+\frac{1}{2}$ ($i = c, h$), and the thermal  population of the oscillator's energy levels, $\bar{n}(x_i)$, has the standard form:
\begin{equation}
\bar{n}\left(\frac{ \omega_i}{T_i}\right)=\frac{1}{ \exp\left[\omega_i/T_i\right]-1}. 
\label{eq:1-auxiliary}
\end{equation}
Below we will also use $\kappa = (2\nu)^{-1} = \tanh(\omega/2T)$.

As expected,  we see from Eq.(\ref{eq:1-work-and-heat}), that here always $R_1 R_2 < 0$ and $Q_1 Q_2 < 0$. The direction of the cycle is determined by the parameter
\begin{equation}
\lambda =  \frac{ \omega_c}{T_c} - \frac{ \omega_h}{T_h}.
\label{eq:1-parameter}
\end{equation}
For $\lambda > 0$, the net work $R = R_1 + R_2 > 0$, and the device works as a heat engine with efficiency
\begin{equation}
\eta = \frac{R}{Q_1} = 1 - \frac{\omega_c}{\omega_h}.
\label{eq:1-eta}
\end{equation}
 For $\lambda < 0$, it becomes a refrigerator with  efficiency
 \begin{equation}
\zeta = \frac{-Q_2}{R} = \frac{\omega_c}{\omega_h - \omega_c}.
\label{eq:1-zeta}
\end{equation} 
It is also straightforward to check that the Clausius inequality, $C \equiv \Delta S - \Delta Q/T \geq 0$, is satisfied on the non-adiabatic stages of the cycle. Indeed, the entropy on the compression stage is
\begin{equation}
S_{AB} =  \ln\left(\nu_c-\frac{1}{2} \right) + \frac{ \omega_c }{T_c}\left(\nu_c+\frac{1}{2} \right) 
= -\ln [2\sinh(\omega_c/2T_c)] + (\omega_c/2T_c) \coth(\omega_c/2T_c),
\label{eq:1-S-AB}
\end{equation}
and for that on the expansion stage, $S_{CD}$, we replace $\left( \omega_c/T_c,\:\nu_c \right)$ with $\left( \omega_h/T_h,\:\nu_h \right)$. From here, using the identity 
\begin{eqnarray}
\nu_h - \nu_c = \frac{\sinh\left(\frac{ \omega_c}{2T_c}-\frac{ \omega_h}{2T_h}\right)}{\sinh\left(\frac{ \omega_c}{2T_c}\right)\sinh\left(\frac{ \omega_h}{2T_h}\right)},
\label{eq:auxiliary-2}
\end{eqnarray}
 and, for positive $x,y$, $$\ln(\sinh x/\sinh y) \leq (x-y) \coth y,$$ we find that indeed 
\begin{eqnarray}
C_{BC} = S_{CD}-S_{AB}-\frac{Q_2}{T_h} \geq 0; \nonumber \\
C_{DA} = S_{AB}-S_{CD}-\frac{Q_1}{T_c} \geq 0.
\label{eq:auxiliary4}
\end{eqnarray} 

An interesting feature of the quantum Otto cycle is that, unlike the Carnot cycle, its efficiency is independent on the temperatures of the heater and cooler (cf. \cite{Quan2007}). The Carnot inequality is not violated though, since precisely at the point where $\eta = \eta_C$, the parameter $\lambda$ goes through zero and   switches sign, turning the heat engine into a refrigerator. However, the independence of the cooling efficiency $\zeta$  on temperature may make a quantum Otto fridge a candidate for the realization of self-cooling qubits. The minimal temperature  such a fridge can cool down the ``cooler" is limited by the condition $\lambda \leq 1$ to 
\begin{equation}
T_c = T_h \frac{\omega_c}{\omega_h}.
\label{eq:1-cooling-limit}
\end{equation}
The minimal temperature that can be achieved using other methods is discussed in \cite{Grajcar2008c}.

\section{Quantum Otto cycle at finite operation speed}

In order to analyze the effects of a finite speed of operation of the Otto engine, it is convenient to use the Wigner function in the basis of coherent states of the harmonic oscillator (e.g., \cite{Gardiner2004}), which was used to investigate  parametric squeezing \cite{Zagoskin2008,Zagoskin2011}. 
It is well known that a sudden change of frequency of a linear oscillator transforms a coherent state into a squeezed state, with the squeezing proportional to the ratio of initial and final frequencies,   while a quasistatic change  does not \cite{Graham1987,Agarwal1991,Janszky1992,Kiss1994}. 

Note that  squeezing  is a unitary ( i.e., reversible ) operation. The von Neumann entropy of a quantum state, $S[\hat{\rho}] = -{\tt tr} \hat{\rho} \ln \hat{\rho}$, is not changed by squeezing. The situation is different for the so-called ``energy entropy" (see, e.g., \cite{Feldmann2006}),
\begin{equation}
S_E[\hat{\rho}] = - \sum_{n=0}^{\infty} p_n \ln p_n \geq S[\hat{\rho}],
\label{eq:SE-definition}
\end{equation}
where $p_n$ are the diagonal components of the density  matrix in the energy representation. Note that $S_E$ and $S$
coincide only if $[\hat{\rho},H] = 0$. The decoherence processes, which eventually eliminate the off-diagonal elements of the density matrix in the energy basis, thus increase the von Neumann entropy and can be associated with the inner friction processes. The energy entropy is therefore a quantitative measure of the inner friction in a quantum system, which is appropriate for the analysis of quantum heat engines operating at finite speed \cite{Feldmann2006}.  The use of $S_E$ has an additional advantage in that it can be directly expressed in terms of the Wigner function.

Consider an arbitrary time dependence of the system frequency $\omega(t)$. This will impose a constant change on the basis of coherent states (which are defined with respect to the Fock space of an oscillator with  instantaneous frequency $\omega(t)$). It is therefore convenient to use the preferred basis (e.g., the set of coherent states in the Fock space with $\omega(0)$). The master equation for the Wigner function $W(\alpha,\: \alpha^*;t)$, with $\alpha,\: \alpha^*$ always referring to this basis, reads \cite{Zagoskin2008,Zagoskin2011}
\begin{equation}
\frac{\partial}{\partial t}W(\alpha, \alpha^*, t) = 2\omega(t) {\tt Im} \left(\alpha^* \frac{\partial}{\partial \alpha^*}\right) W(\alpha, \alpha^*, t) + \frac{\partial \ln \omega(t)}{\partial t} {\tt Re} \left(\alpha \frac{\partial}{\partial \alpha^*}\right) W(\alpha, \alpha^*, t),
\label{eq:4-squeezing-W-complex}
\end{equation}
or 
\begin{equation}
\frac{\partial}{\partial t}W(x, y, t) = \omega(t)   \left(x\frac{\partial}{\partial y} - y\frac{\partial}{\partial x}\right) W(x, y, t) + \frac{1}{2}\frac{\partial \ln \omega(t)}{\partial t}   \left(x \frac{\partial}{\partial x} - y \frac{\partial}{\partial y}\right) W(x, y, t).
\label{eq:4-squeezing-W-real}
\end{equation}
\\

We omitted the diffusion terms, which describe decoherence (including relaxation). Therefore these equations are valid for the adiabatic stages of the cycle (assuming that the system has no inrinsic sources of decoherence).  Equation~(\ref{eq:4-squeezing-W-real}) is a  first-order linear equation and can be solved by the method of characteristics: using the Ansatz $W(x,y,t) \equiv W(x(t), y(t))$ we find from (\ref{eq:4-squeezing-W-real}) the  characteristic equations,
\begin{eqnarray}
\frac{dx}{dt} = \frac{\dot{\omega}}{2\omega} x - \omega y; \:\:
\frac{dy}{dt} =  \omega x - \frac{\dot{\omega}}{2\omega} y.
\label{eq:4-characteristics}
\end{eqnarray}
The evolution of the Wigner function $W(t) \equiv W(x_0(x,y,t), y_0(x,y,t))$  is due to the initial distribution being ``dragged" along the characteristic curves. The total phase volume occupied by the system obviously will not change during such an evolution. As shown in \cite{Zagoskin2008,Zagoskin2011}, in the limit of fast frequency change, when the $\dot{\omega}$-terms dominate, these equations lead to the squeezing of the Wigner function, while in the quasistatic limit they simply describe its rotation as a whole, without disrupting an equilibrium state.  For example, an instantaneous change of the oscillator frequency  $\omega_c \to \omega_h$  will transform the thermal state into a squeezed thermal state (see \cite{Kim1989}, Eq.(4.13)) characterized by the squeezing parameter $s = \omega_h/\omega_c$ :
\begin{equation}
W^{ST}(x,y) = \frac{1}{\pi \bar{n}} \exp\left[ - \frac{1}{\bar{n}}\left(x^2 s + y^2/s\right)\right].
\label{eq:squeezed-therm}
\end{equation}

On the isothermic stages, the r.h.s. of Eq.~(\ref{eq:4-squeezing-W-real}) must also include the diffusive term \cite{Gardiner2004} 
\begin{equation}
\gamma_{c(h)}\left[   \frac{\nu_{c(h)} }{4}\left(\frac{\partial^2}{\partial x^2} + \frac{\partial^2}{\partial y^2}\right) + \frac{\partial}{\partial x} x + \frac{\partial}{\partial y} y\right] W(x,y,t),
\label{eq:1-diffusive}
\end{equation} 
where $\gamma_{c(h)}$ 
is the decay rate 
of the corresponding reservoir. Then the equations cannot be solved analytically and must be dealt with numerically. 

The energy of the system and the squeezing coefficient are expressed through the Wigner function as
\begin{eqnarray}
E = \langle H \rangle =  \omega \int\!\!\int dxdy\: (x^2+y^2)\:W(x,y,t); \label{eq:5-44bis} \\
s(t) =  {\rm max}_{\theta}\left\{ s_{\theta}(t)\right\} \equiv {\rm max}_{\theta} \left\{ \frac{\int\!\!\int dxdy\: (x \cos\theta + y \sin\theta)^2 \:W(x,y,t)}{\int\!\!\int dxdy\:  (y \cos\theta - x \sin\theta)^2 \:W(x,y,t)} \right\}.
\label{eq:5-44}
\end{eqnarray}
In the last expression we take into account that the ``cigar" of the squeezed state rotates in the phase plane $xy$ with frequency $\omega(t)$.

In order to calculate the energy entropy (\ref{eq:SE-definition}) we  use the  expression for the Wigner function of a Fock state (e.g., Eq.(4.4.91) \cite{Gardiner2004}):
\begin{equation}
|n\rangle\langle n| \leftrightarrow W^F_n (\alpha,\alpha^*) = \frac{2(-1)^n}{\pi} e^{-2|\alpha|^2} L_n(4|\alpha|^2).
\label{eq:WFn}
\end{equation}
Here $L_n(x)$ is the Laguerre polynomial. 
Finding the diagonal part of the density matrix in energy representation is now straightforward. It can only depend on the angle-averaged Wigner function
\begin{equation}
\overline{W}(|\alpha|^2) = \frac{1}{2\pi}\int_0^{2\pi}\! d\theta\; W(|\alpha|e^{i\theta},|\alpha|e^{-i\theta}).
\label{eq:W-averaged}
\end{equation}
Expanding a well-behaved function of variable $\eta = 4|\alpha|^2$, $f(\eta) = \overline{W}(\eta/4)\exp[\eta/2]$, in the Laguerre series,
\begin{equation}
f(\eta) = \sum_{n=0}^{\infty} A_n L_n(\eta),\:\: A_n = \int_0^{\infty}\! d\eta \,f(\eta) \,L_n(\eta) \,e^{-\eta},
\label{eq:f-exp}
\end{equation}
we see, that 
\begin{equation}
\overline{W}(|\alpha|^2) = \sum_n^{\infty} \frac{\pi(-1)^n}{2} A_n W^F_n(|\alpha|^2).
\label{eq:W-exp}
\end{equation}
Therefore, the diagonal matrix elements of the density matrix in the energy representation are $p_n = \pi(-1)^n A_n/2$. Writing 
\begin{equation}
d\alpha d\alpha^* = |\alpha| d|\alpha| d\theta = \frac{1}{8} d\eta d\theta
\label{eq:differential}
\end{equation}
and using (\ref{eq:f-exp}) and (\ref{eq:W-averaged}), we obtain
\begin{equation}
p_n = 2 (-1)^n \int d\alpha d\alpha^*\, e^{-2|\alpha|^2} L_n(4|\alpha|^2)\, W(\alpha,\alpha^*).
\label{eq:pn-general}
\end{equation}
In terms of $x = {\tt Re}\alpha, y = {\tt Im}\alpha$, this is
\begin{equation}
p_n = 2 (-1)^n \int\!\!\int dx dy \,e^{-2(x^2+y^2)}\, L_n(4(x^2+y^2)) \,W(x,y).
\label{eq:pn-xy}
\end{equation}
These expressions allow a direct calculation of the diagonal elements of the density matrix, $\{p_n\}_{n=0..\infty}$, and with that, of $S_E[\hat{\rho}]$. 

One can also introduce the quasiclassical entropy, $S_{qc}[\hat{\rho};\chi],$ utilizing the fact that the Wigner function $W(p,q)$ reduces to a classical probability distribution after being averaged over the scale of $\Delta p \Delta q \geq h$ (in our case, $\Delta x \Delta y \geq 1$)\cite{Ropke1987}:
\begin{eqnarray}
S_{qc}[\hat{\rho};\chi] = - \int\!\!\int dxdy \left( \left[\int\!\!\int dx'dy' W(x',y')\chi(x-x',y-y')\right]\times \right. \nonumber\\
\left.  \ln \left[\int\!\!\int dx'dy' W(x',y')\chi(x-x',y-y')\right]\right).
\label{eq:S_qc}
\end{eqnarray}
Here $\chi(x,y)$ is a normalized sampling function peaked at zero, with a support dimension exceeding unity (e.g., $\chi = (\pi d^2)^{-1} \exp[-(x^2+y^2)/d^2], \: d > 1/2$).

As a relevant example, let us calculate the energy entropy of the squeezed thermal state. From Eq.(\ref{eq:squeezed-therm}) we can write
\begin{equation}
p_n = 2 (-1)^n \frac{2\kappa}{\pi} \int_0^{2\pi}d\theta \int_0^{\infty}dr\: r e^{-2r^2 - 2r^2\kappa (s\cos^2\theta+(1/s)\sin^2\theta)}L_n(4r^2),
\label{eq:pn-intermediate-1}
\end{equation}
which can be rewritten as
\begin{equation}
p_n = (-1)^n \frac{\kappa}{2\pi} \int_0^{2\pi}d\theta \int_0^{\infty}d\eta \exp\left\{-\frac{\eta}{2}\left[1 +\frac{\kappa}{s} + \kappa(s-1/s)\cos^2\theta\right]\right\} L_n(\eta).
\label{eq:pn-intermediate-2}
\end{equation}
Using the formula 
\begin{equation}
\int_0^{\infty} d\eta L_n(\eta) e^{-Q\eta} = \frac{(Q-1)^n}{Q^{n+1}} = (-1)^n \sum_{q=0}^{n} C_n^q (-1)^q Q^{q-n-1},
\label{eq:L-int}
\end{equation}
we see, that
\begin{equation}
p_n = \frac{\kappa}{2\pi}\int_0^{2\pi} d\theta  \frac{\left\{\frac{1}{2}\left[1 - \frac{\kappa}{s} - \kappa(s-1/s)\cos^2\theta\right]\right\}^n}{\left\{ \frac{1}{2}\left[1 +\frac{\kappa}{s} + \kappa(s-1/s)\cos^2\theta\right]\right\}^{n+1}},
\label{eq:pn-intermediate-3-bis}
\end{equation}
or
\begin{equation}
p_n = \frac{\kappa}{2\pi}\int_0^{2\pi} d\theta  \sum_{q=0}^{n} C_n^q (-1)^q \left\{\frac{1}{2}\left[1 +\frac{\kappa}{s} + \kappa(s-1/s)\cos^2\theta\right]\right\}^{q-n-1}
\label{eq:pn-intermediate-3}
\end{equation}
Using the table integral (for $c,d >0$)
$$
\int_0^{2\pi} \frac{d\theta}{(c + d \cos^2\theta)^m} = \frac{2\pi}{c^m} \left.\right._2{\rm F}_1\left(\frac{1}{2},m;1;-\frac{d}{c}\right),
$$
we finally obtain from Eq. (\ref{eq:pn-intermediate-3})
\begin{equation}
p_n = \kappa \sum_{q=0}^n C_n^q (-1)^q \left(\frac{2}{1+\frac{\kappa}{s}}\right)^{n+1-q} \left.\right._2{\rm F}_1\left(\frac{1}{2},n+1-q;1;-\frac{\kappa(s-1/s)}{1+\kappa/s}\right)
\label{eq:pn-final-thermal}
\end{equation}
and the corresponding expression for $S_E$. 

To check this, consider the non-squeezed case, $s=1$. Then the hypergeometric function in (\ref{eq:pn-final-thermal}) is $\left.\right._2{\rm F}_1\left(\dots,\dots;\dots;0\right) = 1$, and the expression reduces to 
$$
\frac{2\kappa}{(1+\kappa)^{n+1}} \left[2 - (1+\kappa)\right]^n = \frac{2\kappa}{1+\kappa} \left(\frac{1-\kappa}{1+\kappa}\right)^n.
$$ 
Substituting $\kappa = \tanh(\omega/2T)$, we see that indeed the populations reduce to their equilibrium values,
\begin{equation}
p_n = (1-e^{-\omega/T})e^{-n\omega/T} \equiv p_n^{\rm eq}.
\label{eq:equilibrium}
\end{equation}

The expression (\ref{eq:pn-final-thermal}), while exact, is not very illuminating. A useful approximation for small squeezing $\left[\kappa\;(s-1) \ll 1\right]$ can be obtained directly from Eq.(\ref{eq:pn-intermediate-3-bis}) rewritten as
\begin{equation}
p_n = \left\{\frac{2\kappa}{\kappa+1} \left[\frac{1-\kappa}{1+\kappa}\right]^n \right\}
\frac{1}{2\pi}\int_0^{2\pi} \!\!\!d\theta\;  
\frac{\left\{1 + \frac{\kappa}{1-\kappa}\left[1-1/s - (s-1/s)\cos^2\theta\right]\right\}^n}{\left\{1 - \frac{\kappa}{1-\kappa}\left[1-1/s - (s-1/s)\cos^2\theta\right]\right\}^{n+1}}.
\label{eq:pn-intermediate-3-bis-2}
\end{equation}
The expression in brackets is simply $p_n^{\rm eq}$, the equilibrium population given by Eq.~(\ref{eq:equilibrium}).

For small values of $\kappa(s-1)$,  powers in the integral can be replaced by exponents,
\begin{equation}
p_n \approx p_n^{\rm eq}
\frac{1}{2\pi}\int_0^{2\pi} d\theta  
\exp\left[n\frac{\kappa(s-1)}{1-\kappa}\left(1-2\cos^2\theta\right)\right]\exp\left[(n+1)\frac{\kappa(s-1)}{1+\kappa}\left(1-2\cos^2\theta\right)\right]. 
\label{eq:pn-intermediate-3-bis-4}
\end{equation}
Since 
$$
\int_0^{\pi} dx \: e^{z \cos x} = \pi I_0(z) = \pi I_0(-z),
$$
we eventually find 
\begin{equation}
p_n \approx p_n^{\rm eq} I_0\left[\kappa(s-1)\left(\frac{n}{1-\kappa}+\frac{n+1}{1+\kappa}\right)\right] \approx p_n^{\rm eq} I_0\left[\frac{\omega (1-s)}{T}(n+1/2)\right]
\label{eq:p-approx}
\end{equation}
(the last simplification works for $\omega/T \ll 1$). Either approximation in (\ref{eq:p-approx})  satisfies the normalization condition: $\sum_n p_n \approx 1$. 

\begin{figure}%
\includegraphics[width=3.2in]{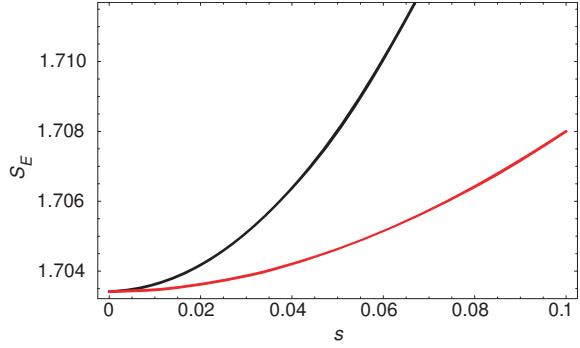}
\caption{(Color online.) Energy entropy $S_E = - \sum_n p_n \ln p_n$ of a squeezed thermal state as a function of the squeezing parameter $s$ for the ratio $\omega/T = 0.5$. The populations of Fock states are given by the exact expressions (\ref{eq:pn-final-thermal}) (red, lower curve) and the approximation (\ref{eq:p-approx}) (black, upper curve).  }%
\label{figAAA}%
\end{figure}

Squeezing depopulates low-energy states and populates the high-energy ones. For small squeezing, the change in energy is
\begin{equation}
\delta\! E = \sum_n n \; \omega\;  \delta p_n \sim \; (s-1)^2
\label{eq:energy-change}
\end{equation}
[since $I_0(z) = 1 + z^2/4 + \dots$]. 
The effect on the energy entropy becomes
\begin{equation}
\delta S_E \approx -\sum_n (\delta p_n)  \ln p_n \approx -\sum_n (\delta p_n)  \ln p_n^{\rm eq} \; \sim \; (s-1)^2.
\label{eq:approx-SE-change}
\end{equation}
This is actually the case, as can be seen in Fig.~\ref{figAAA}. This increase in the energy and the entropy production (and the resulting increase in the heat transfer to the reservoirs) will reduce the efficiency of the Otto cycle compared to the quasistatic case.

\section{Simulations}

The equation for the Wigner function $W(t)$ has both a drift term (\ref{eq:4-squeezing-W-real}) and a diffusion term (\ref{eq:1-diffusive}) and formally coincides with the Fokker-Planck equation 
\begin{equation}
\partial P/\partial t = \nabla\left[P\nabla U(x,y,t)\right] - D(t)\;\nabla^2 P
\label{eq:Focker-Planck}
\end{equation}
where $P$ is the probability density of a Brownian particle
to be near the point $(x,y)$, $U(x,y,t)$ is the time-dependent potential describing deterministic forces acting on the particle, while $D$ is the diffusion coefficient. It is well known that the Fokker-Planck equation is equivalent to the stochastic Langevin equation $(\dot x, \dot y)=-\nabla U
+\xi$, with stochastic force $\xi$ satisfying the following conditions \cite{Savelev2004}: $$\langle \xi\rangle=0; \:\:\: \langle \xi(t)\xi(0)\rangle=D\delta(t).$$ Therefore,   the equation for the Wigner function $W(t)$ should be equivalent to a set of stochastic equations:
\begin{eqnarray}
\dot x=\omega(t)y-\frac{\dot\omega}{2\omega}x-\frac{\gamma_{c(h)}(t)}{2}x+\xi_x\nonumber \\
\dot y=-\omega(t)x+\frac{\dot \omega}{2\omega}y-\frac{\gamma_{c(h)}(t)}{2}+\xi_y,
\label{langevin}
\end{eqnarray}      
where the effective diffusion constant is defined as 
\begin{equation}
D=(1/4)\gamma_{c(h)}(t)\left(1+2/\left\{\exp\left[\omega(t)/T_{c(h)}(t)\right]-1 \right\}\right).
\label{eq:D}
\end{equation}

\begin{figure}%
\includegraphics[width= 3.2 in]{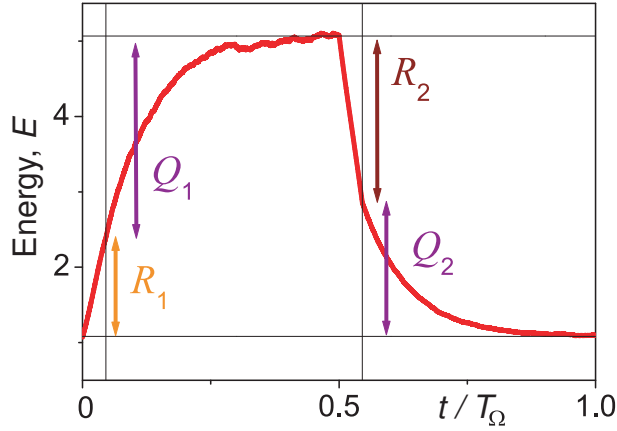}
\caption{(Color online.) Energy $E$ of the system as a function of the reduced time, $t/T_{\Omega},$ during the heat engine cycle  ($|R_1| < |R_2|$). }%
\label{fig-ENERGY}%
\end{figure}

\begin{figure}%
\includegraphics[width=3.2 in]{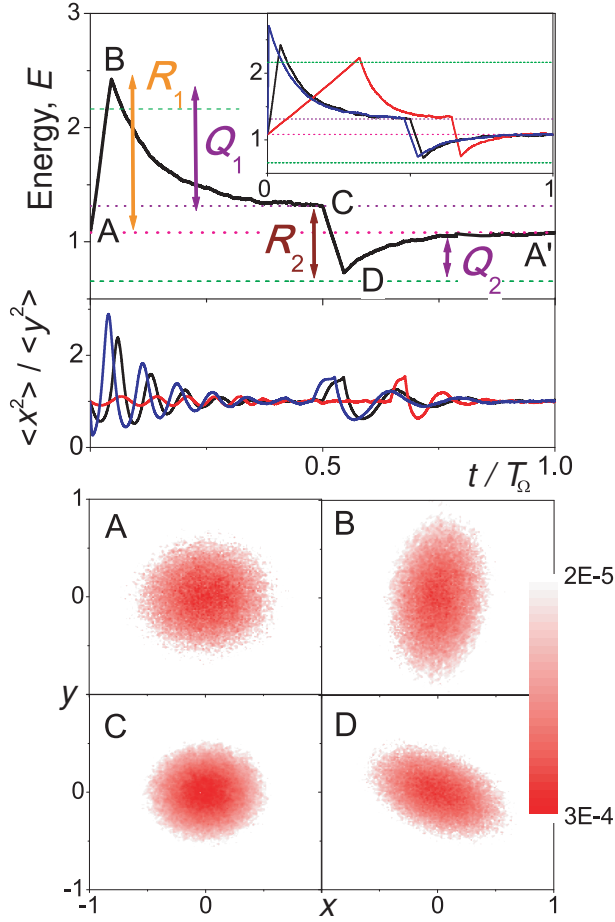}
\caption{(Color online.) (Top) Energy $E$ of the system as a function of time during the heat pump cycle $|R_1| > |R_2|$. As the rate of adiabatic expansion decreases, the system approaches the operation of a quasistatic Otto cycle (inset). (Middle) The ratio $(\langle x^2 \rangle/\langle y^2 \rangle)$ of quadrature dispersion rates during the cycle, for the same expansion rates as above. Oscillations are due to the rotation of the Wigner function in the phase plane with the instantaneous eigenfrequency of the oscillator. (Bottom) Wigner function in the phase plane at different points (A, B, C, D) of the Otto cycle (also shown in Fig.~1).}%
\label{fig-ALL}%
\end{figure}

Using the Langevin equations (\ref{langevin}) we can easily simulate the evolution of the Wigner function. Starting at $t=0$ from the thermal
Wigner function (\ref{eq:squeezed-therm}), we randomly spread $N$ Brownian particles according to this Gaussian probability density. Then we numerically monitor the evolution of each of these Brownian particles with increasing $t$. The obtained distributions of particles 
at any $t$ allow  us to numerically estimate the time dependence of the Wigner function and calculate the energy of the system 
$E= \omega(t)\langle x^2+y^2\rangle$ and the squeezing coeffecient $\beta=\langle x^2\rangle/\langle y^2\rangle$, when arbitrarily changing $\omega$ with time. The results are shown in Figs. \ref{fig-ENERGY}, \ref{fig-ALL}.

\begin{figure}%
\includegraphics[width=3.2 in]{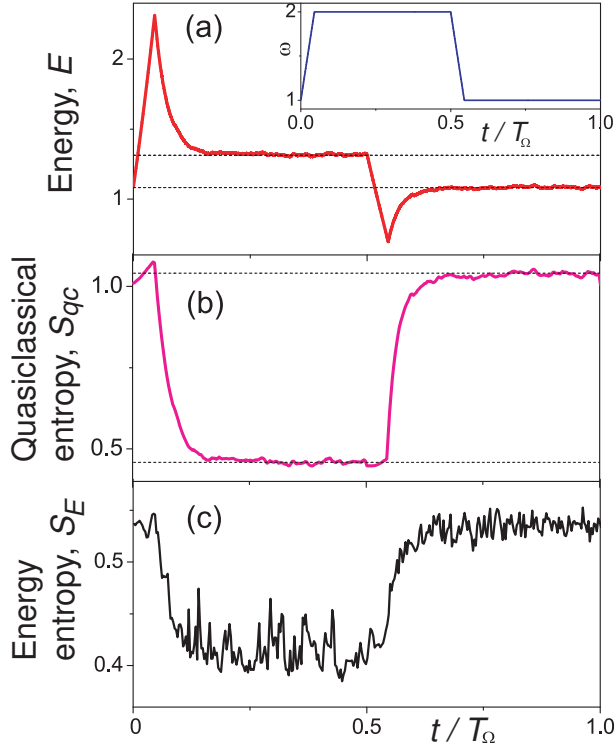}%
\caption{ (Color online.) Energy (a) and entropy (b,c) of the quantum Otto engine during the cycle shown in Fig.~1(b). Inset: time dependence of the oscillator frequency, $\omega$. }%
\label{fig-entropy}%
\end{figure}

The {\it energy entropy}  of the system, $S_E(t)$, is calculated substituting (\ref{eq:pn-xy}) in Eq.(\ref{eq:SE-definition}). We also calculate the {\it quasiclassical entropy}, $S_{qc}(t)$, from Eq.(\ref{eq:S_qc}). In the latter case instead of using the sampling function $\chi$ we directly counted the number of representing particles within cells $[x_i+\Delta x, y_i+\Delta y]$ (i.e., using a grid). The results are shown in Fig.~\ref{fig-entropy}.

The squeezing due to the finite expansion/compression rate is the only source of inefficiency in our model system. Compared to the quasistatic case, it leads to increased entropy and energy at the points B and D. Thus, e.g., in the heat engine regime, decreasing the net work performed by the system and increasing the net heat transfer to the cold reservoir. 

From Fig.  \ref{fig-ALL} we see that the performance of the quantum Otto engine fast approaches its quasistatic limit as the squeezing of the quantum state of the working body (tunable oscillator) is decreased. This is consistent with the effects of squeezing being quadratic in the squeezing parameter, though the roughness of calculations does not allow to confirm the exact functional shape of this dependence.

\section{Conclusions}

We have analyzed a simple model of a quantum Otto engine, which can be realized based on, e.g.,  Josephson devices. The roles of the working body and hot and cold reservoirs are played by oscillators (with tunable or fixed frequencies, respectively). We have shown 
that, depending on the relation between temperature and resonant frequencies of the reservoirs, the system can work either as a heat engine or a heat pump. Using the method of Wigner functions, we found that the source of inefficiency of this device is in the squeezing of the quantum state of the working-body oscillator. In particular, we found an explicit expression for the energy entropy of a squeezed thermal state. Though inevitable for any finite speed of operation, the effect of small squeezing $s \approx 1$ is only proportional to $(1-s)^2$.

Comparing the efficiencies of the quantum and thermal engines, one can say that a quantum Otto engine may operate with a high efficiency even at low temperatures, unlike the classical case. The losses   can be minimized by reducing the squeezing of the working body. It may be worth considering incorporating quantum Otto engines in superconducting qubit registers as additional coolers.

 \section{Acknowledgements}

AZ and SS acknowledge that this publication was made possible through the support of a grant from the John Templeton Foundation; the opinions expressed in this publication are those of the authors and do not necessarily reflect the views of the John Templeton Foundation. 
FN is partially supported by the ARO, NSF grant No. 0726909, JSPS-RFBR contract No.12-02-92100, Grant-in-Aid for Scientific Research (S), MEXT Kakenhi on Quantum Cybernetics, and the JSPS via its FIRST program. FVK acknowledges the support of the ESF network-program Arrays of Quantum Dots and Josephson Junctions (AQDJJ) and EPSRC grant EP/I01490X/1.


\end{document}